\documentclass[twocolumn]{el-author}

\date{31th August 2019}

\usepackage{changepage}
\usepackage{graphicx}
\usepackage{enumerate}
\usepackage{amsmath}
\usepackage{diagbox}
\usepackage[justification=centering]{caption}
\usepackage{algorithmicx}
\usepackage{algorithm}
\usepackage{algpseudocode}
\usepackage{amsfonts}
\usepackage{amssymb}
\usepackage{float}
\usepackage{color}
\usepackage{caption}
\DeclareCaptionLabelSeparator{twospace}{\ ~}   
\usepackage[misc]{ifsym}
\newcommand{\beq}{\begin{equation}} \newcommand{\eeq}{\end{equation}}
\newcommand{\ju}{{j\mkern1mu}}

\begin{document}

\title{Transmitting Extra Bits by Rotating Signal Constellations}

\author{Jiachen Sun, Hao Liu and Xiao Ma\textsuperscript{\Letter}
}

\abstract{In this letter, we propose a novel LDPC coding scheme to transmit extra bits aided by rotated signal constellations without any additional cost in transmission power or bandwidth. In the proposed scheme, the LDPC coded data are modulated by a rotated two-dimensional signal constellation, in which the rotation angle is specified by the given extra bits. At the receiver, the rotation angle is estimated with the aid of the statistical learning of the
syndrome of the LDPC code. After recovering the rotation angle, the coded payload data can be decoded by the LDPC decoder. The simulation results show that, for an LDPC code of length 2304, up to four extra bits can be transmitted with negligible influence on the reliability of the LDPC coded data.}

\maketitle

\section{Introduction}
In many wireless communication systems,
a few extra bits are usually required to be transmitted as control signallings along with coded payload data. For instance, in LTE~\cite{dahlman20134g}, an ACK/NCAK bit is used for the HARQ scheme to indicate whether a retransmission of a packet is required or not. Typically, the extra bits and the payload data are transmitted separately~\cite{larsson2012piggybacking} either because these extra bits usually require higher reliability compared with the payload data, or because the receiver only wants to detect the extra bits without waiting decoding the whole block of data. However, the conventional separated-transmission scheme inevitably leads to an extra consumption of power or bandwidth. So far several methods in the literature have been proposed to tackle this challenge. In~\cite{larsson2012piggybacking}, one additional bit is piggybacked on the payload data encoded by a linear channel code. The basic idea is to use the additional bit to specify which of two linear codes is used for coding. In~\cite{hong2015additional}, a rotated QPSK modulation scheme was proposed to carry one extra bit, while approaching the same performance as the original QPSK in the relatively high signal-to-noise ratio (SNR) region. In~\cite{yan2011novel}, the authors developed a new space-time block code transmission scheme, in which one additional bit is used to represent one of two labeled constellations and thus can be transmitted without bandwidth expansion. Recently, the authors in~\cite{cai2019pack} proposed to use superposition to pack additional bits in LDPC coded data at the expense of a slight increase in decoding complexity.

In this work, a novel low-density parity-check~(LDPC) coding scheme is proposed to transmit more than one extra bit with the aid of rotated signal constellations. The underlying principle is to map the coded payload data into a \emph{rotated} two-dimensional signal constellation with a rotation angle specified by the given extra bits. At the receiver, the estimation of rotation angle can be solved by the brute-force search since the number of candidate solutions is typically small. Due to the inherent symmetry of some fixed signal constellation, there may exist more than one rotation angle achieving the same level of the objective function, which is the so-called permutation ambiguity. To solve this problem, we propose to select the estimated rotation angle from a list of candidates by the statistical learning of the syndrome of the LDPC code. After recovering the rotation angle, the coded payload data can be decoded by the LDPC decoder. The simulation results show that up to four extra bits can be transmitted with negligible influence on the reliability of the LDPC coded data.

\begin{figure}[h]
  \centering
  \includegraphics[width=8.5cm]{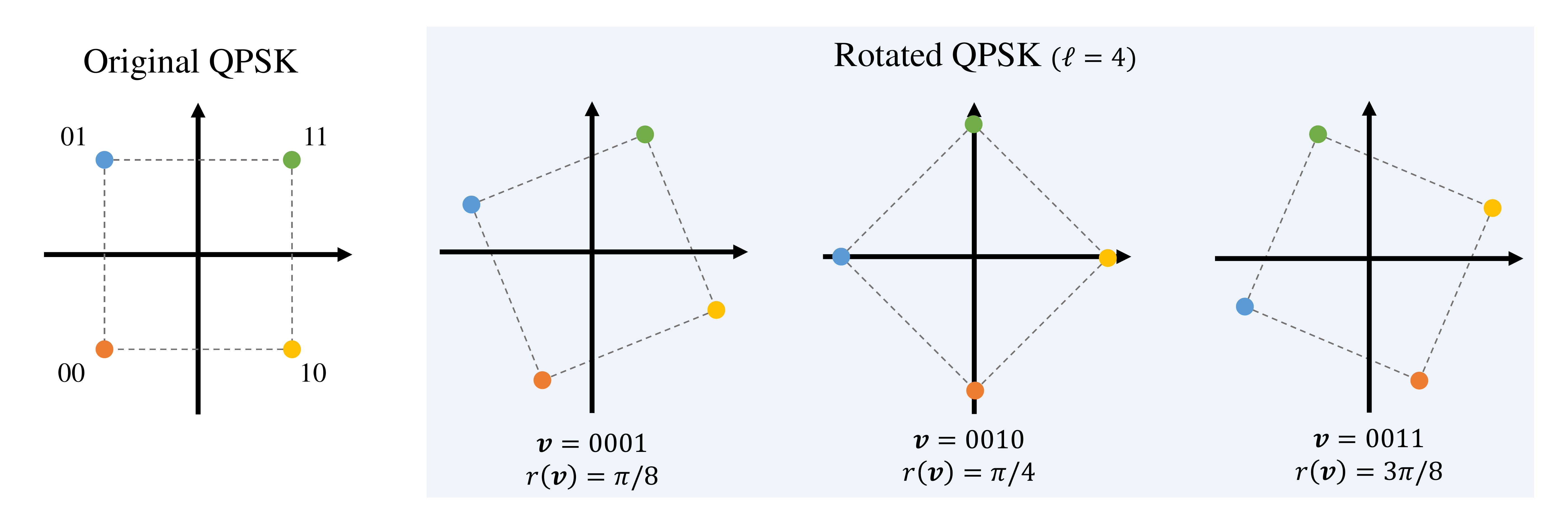}\\
  \caption{Rotated QPSK signal constellations.}\label{FIG0}
\end{figure}

\section{Transmission Scheme}
 Let $\mathcal{C}[N, K]$ be a binary LDPC code with length $N$ and dimension $K$. Let $\mathcal{S} = \{s_0,s_1,...,s_{2^m-1} \} \subset \mathbb{C}$ be a two-dimensional signal constellation with $|\mathcal{S}| = 2^m$. Without loss of generality\footnote{If necessary, redundant zeros can be appended to a codeword.}, we assume that $N = mn$ for some positive $n$. Let $\textbf{\emph{u}} = (u_0,u_1,...,u_{K-1})\in \mathbb{F}_2^K$ be the information sequence~(payload data) and $\textbf{\emph{v}} = (v_0, v_1,...,v_{\ell}) \in \mathbb{F}_2^{\ell}$ be the extra bit sequence with length $\ell$.  The payload data is encoded by the LDPC code $\mathcal{C}[N, K]$, resulting in a codeword sequence $\textbf{\emph{c}} = (c_0,c_1,...,c_{N-1})$.
 The coded payload data $\textbf{\emph{c}}$ is partitioned into $n$ groups, each of which has $m$ bits and is mapped into a constellation point in $\mathcal{S}$, resulting a sequence of modulated signals $\textbf{\emph{x}} = (x_0,x_1,...,x_{n-1})$, where $x_t \in \mathcal{S}$ for $t=0,1,\dots,n-1$.

\begin{figure}[h]
  \centering
  \includegraphics[width=9cm]{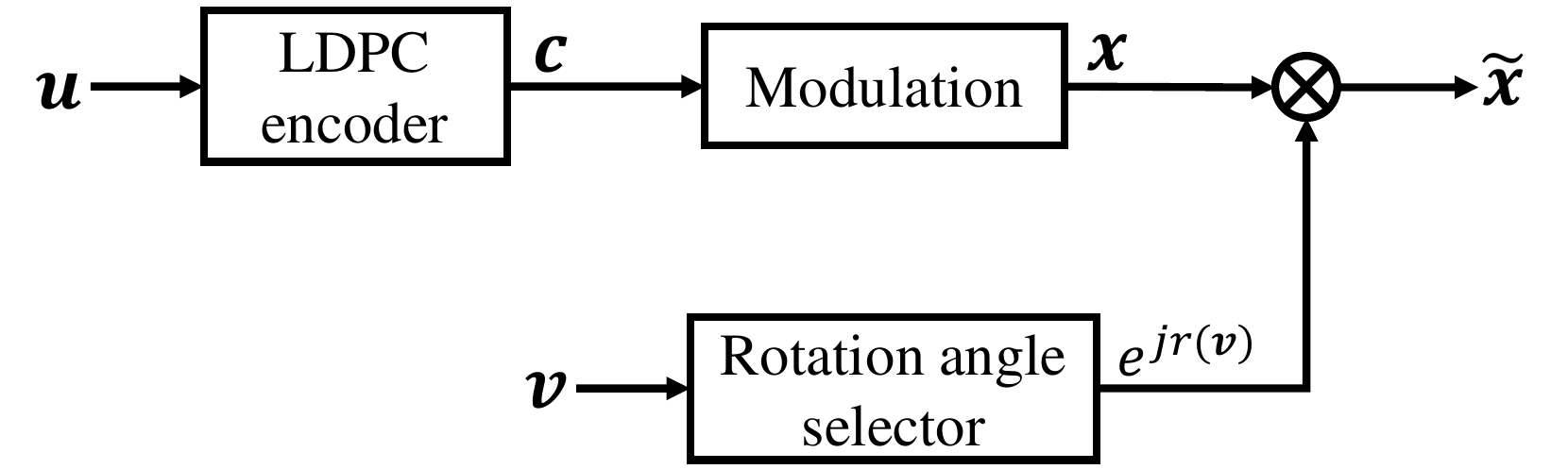}\\
  \caption{The encoder of the proposed scheme.}\label{FIG1}
\end{figure}

To transmit extra bit sequence $\textbf{\emph{v}}$, the modulated signals $\textbf{\emph{x}}$ are then rotated component-wise by a rotation angle $r(\textbf{\emph{v}}) = {2\pi d(\textbf{\emph{v}})}/{2^{\ell}} \in [0, 2\pi)$, where $d(\textbf{\emph{v}})$ is the decimal representation of $\textbf{\emph{v}}$~(see Fig. \ref{FIG0} for an example of three different rotated QPSK constellations with $\ell=4$). Algorithm~1 presents the encoding algorithm of the proposed scheme~(see Fig. \ref{FIG1} for reference).

\begin{algorithm}[h]\caption{Encoder of the proposed scheme}
\begin{adjustwidth}{0.3cm}{}
\begin{enumerate}
\item Encode the information sequence $\textbf{\emph{u}} = (u_0,u_1,...,u_{K-1})$ with the encoder of the LDPC code $\mathcal{C}[N, K]$ to obtain the codeword $\textbf{\emph{c}} = (c_0,c_1,...,c_{N-1})$.
\item Modulate the codeword using the signal constellation $\mathcal{S}$, resulting in a sequence of modulated signals $\textbf{\emph{x}} = (x_0,x_1,...,x_{n-1})$.
\item Rotate component-wise the modulated signals $\textbf{\emph{x}}$ by $r(\textbf{\emph{v}}) = {2\pi d(\textbf{\emph{v}})}/{2^{\ell}} $, where $d(\textbf{\emph{v}})$ is the decimal representation of the extra bit sequence $\textbf{\emph{v}}$, resulting in the transmitted sequence $\widetilde{\textbf{\emph{x}}}$. That is, $\widetilde{\textbf{\emph{x}}} =\exp(\ju r(\textbf{\emph{v}}))\textbf{\emph{x}}$, where $j = \sqrt{-1}$.
\end{enumerate}
\end{adjustwidth}
\end{algorithm}
\section{Decoding Algorithm}
The rotated modulated sequence $\widetilde{\textbf{\emph{x}}}$ is transmitted over an additive white Gaussian noise (AWGN) channel, resulting in a received sequence $\textbf{\emph{y}} = \widetilde{\textbf{\emph{x}}} + \textbf{\emph{w}}$, where the noise samples $w_t$'s are independent and identically distributed (i.i.d)
Gaussian random variables with distribution $\mathcal{CN}(0,\sigma^2)$. The received sequence is denoted by $\textbf{\emph{y}}$. Given $\textbf{\emph{y}}$, the decoder firstly attempts to decode the additional bit sequence $\textbf{\emph{v}}$ through estimating the rotation angle $r(\textbf{\emph{v}})$ of the signal constellation. After recovering the rotation angle, the effect of the additional bit sequence on the received sequence can be removed. Then the LDPC decoder is implemented to decode the information sequence $\textbf{\emph{u}}$.
Obviously, the estimation of the rotation angle $r(\textbf{\emph{v}})$ is critical to reliably decode the payload data.
In the following, we describe in detail a two-stage approach based on the brute-force search and the statistical learning to estimate $r(\textbf{\emph{v}})$ .

Upon receiving $\textbf{\emph{y}} = (y_0,y_1,...,y_{n-1})$, the maximum likelihood~(ML) estimation of the rotation angle is to find $\theta \in [0,2\pi)$ that maximizes the following objective function
\begin{equation}\label{eqn_1}
{F}(\theta) =  \sum_{t=0}^{n-1} {\rm log} \sum_{i=0}^{2^m-1}\exp(-\frac{\| y_t - s_i\exp(j\theta) \|^2}{\sigma^2}).
\end{equation}
Since $\ell$ is typically small, (1) can be solved, say, by the brute-force search which examines all possible $2^{\ell}$ valid rotation angles and finds an angle $\widehat{\theta}$ that maximizes $F(\theta)$ in (1).
Notice that, there may exist more than one rotation angle achieving the same level of objective function $F(\theta)$, which is the so-called permutation ambiguity.
For instance, for QPSK, two rotation angles $\pi$ and $\pi/2$ can achieve the same level of $F(\theta)$ but with different mappings between the transmitted bits and the constellation points.
This is due to the inherent symmetry of some constellations.
To solve this problem, we form a list of rotation angle candidates, denoted by $\Theta = \{\theta_1,\theta_2,...,\theta_I \}$, such that all candidates $\theta_i \in \Theta$  achieve the same maximal level of the objective function $F(\theta)$ in (1). The question is how to identify efficiently the right rotation angle from the list of candidates $\Theta$ provided that the correct one is contained in the list\footnote{To ensure with high probability as required that the correct rotation angle is contained in the list, we can further enlarge $\Theta$  by including all valid angles $\theta$ with $F(\theta) \geq T$ for some pre-designed threshold $T$.}. This identification can be done by learning the statistical behaviors of the syndromes of the LDPC codes.

For each candidate $\theta_i \in \Theta$ for $i = 1,2,...,I$, we firstly rotate  component-wise $\textbf{\emph{y}}$ by $-\theta_i$ to obtain an estimated modulated signals $\widetilde{\textbf{\emph{y}}}^{(i)}$, i.e., $\widetilde{\textbf{\emph{y}}}^{(i)} = \exp(-j \theta_i)\textbf{\emph{y}}$. Then we make hard decision on $\widetilde{\textbf{\emph{y}}}^{(i)}$ to obtain the hard-decision sequence $\widetilde{\textbf{\emph{c}}}^{(i)}$. Let $\mathbf{H}$ be the parity-check matrix of $\mathcal{C}[N,K]$ and  $W(\widetilde{\textbf{\emph{c}}}^{(i)}\mathbf{H}^{\mathrm{T}})$ be the Hamming weight of $\widetilde{\textbf{\emph{c}}}^{(i)}\mathbf{H}^{\mathrm{T}}$. Equivalently, $W(\widetilde{\textbf{\emph{c}}}^{(i)}\mathbf{H}^{\mathrm{T}})$ denotes the number of unsatisfied parity-check equations. Intuitively but importantly, the typical value of $W(\widetilde{\textbf{\emph{c}}}^{(i)}\mathbf{H}^{\mathrm{T}})$ with the correct rotation angle $\theta_i$ is different from that of $W(\widetilde{\textbf{\emph{c}}}^{(i)}{\mathbf{H}}^{\mathrm{T}})$ with the erroneous $\theta_i$, as verified by the following example.

\begin{figure}[h]
  \centering
  \includegraphics[width=6.5cm]{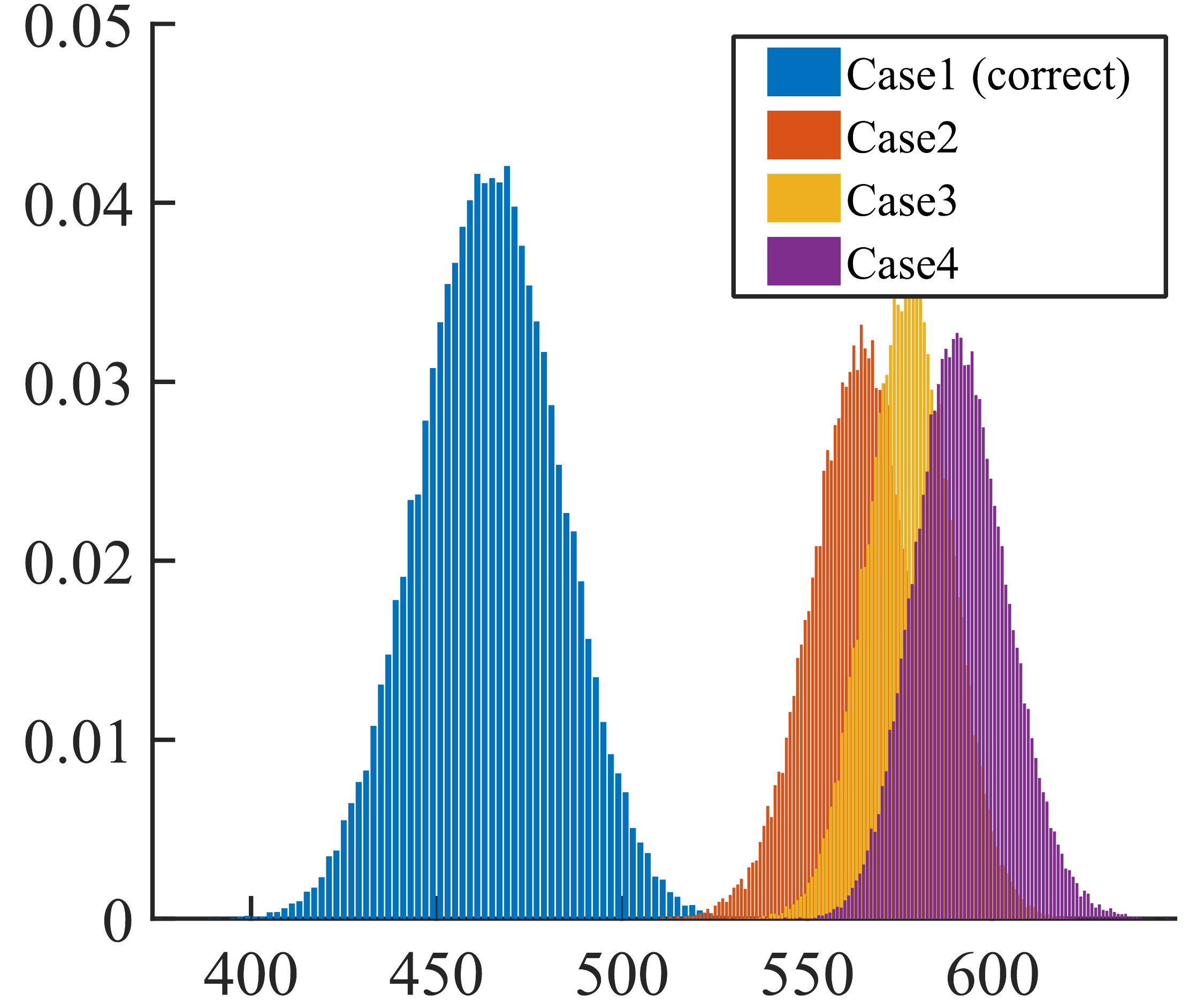}\\
  \caption{Statistical behavior of $W(\widehat{\textbf{\emph{c}}}{\mathbf{H}}^{\mathrm{T}})$.}\label{FIG2}
\end{figure}

\textbf{\emph{Example 1:}} In this example, we select a rate-$1/2$ $(3,6)$-regular LDPC code of length 2304 to present the typical values of $W(\widetilde{\textbf{\emph{c}}}^{(i)}{\mathbf{H}}^{\mathrm{T}})$.  We consider the Gray 16QAM and set the number of extra bits $\ell$ to $3$. Due to the symmetry of the 16QAM constellation, there exist four rotation angles that maximize the objective function $F$, but only one among which is the correct one. Fig. \ref{FIG2} illustrates the histogram of $W(\widetilde{\textbf{\emph{c}}}^{(i)}{\mathbf{H}}^{\mathrm{T}})$ with the correct rotation angle and other three erroneous ones. As expected, $W(\widetilde{\textbf{\emph{c}}}^{(i)}{\mathbf{H}}^{\mathrm{T}})$ is likely to be small if the angle is correct, while $W(\widetilde{\textbf{\emph{c}}}^{(i)}{\mathbf{H}}^{\mathrm{T}})$ is likely to be large if the angle is erroneous.

The above observations provide us with a simple approach to distinguish the correct rotation angle from the erroneous ones. The rotation angle corresponding to the least number of unsatisfied parity-check equations is treated to be the correct candidate. To summarize, the decoder of the proposed scheme is described in Algorithm~2.

\begin{algorithm}\caption{Decoder of the proposed scheme}
\begin{adjustwidth}{0.3cm}{}
\begin{enumerate}
\item Implement the brute-force search to find a valid rotation angle $\widehat{\theta}$ that maximizes the objective function $F(\theta)$ in (1).
\item For the symmetric constellation, find all valid angles $\widehat{\theta}$ that maximize $F(\theta)$ to obtain a list of candidates, denoted by $\Theta = \{\theta_1,\theta_2,...,\theta_I \}$.
\item For each $\theta_i \in \Theta$ for $i=1,2,...,I$, compute $\widetilde{\textbf{\emph{y}}}^{(i)} = \exp(-\ju \theta_i)\textbf{\emph{y}}$.
\item Make hard-decision on $\widetilde{\textbf{\emph{y}}}^{(i)}$ to obtain $\widetilde{\textbf{\emph{c}}}^{(i)}$.

\item The estimated rotation angle is given by
$\widehat{\theta}$ with $\min_{1\leq i\leq I} \{W(\widetilde{\textbf{\emph{c}}}^{(i)}{\mathbf{H}}^{\mathrm{T}})\}$.
Given $\widehat{\theta}$, the estimated extra bits $\widehat{\textbf{\emph{v}}}$ can be determined accordingly.

\item Remove the influence of $\widehat{\textbf{\emph{v}}}$ on ${\textbf{\emph{y}}}$ by computing $\widetilde{\textbf{\emph{y}}} = \exp(-j \widehat{\theta}) {\textbf{\emph{y}}}$.

\item Input $\widetilde{\textbf{\emph{y}}}$ into the LDPC decoder to obtain the estimated information sequence $\widehat{\textbf{\emph{u}}}$.
\end{enumerate}
\end{adjustwidth}
\end{algorithm}

\section{Complexity Analysis}
Notice that, the proposed scheme does not require any bandwidth expansion or transmission energy consumption to transmit extra bits. The complexity of additional operations with the proposed scheme is caused by the brute-force search as well as the LDPC-aided check of rotation angle candidates. For the brute-force search, the complexity is about $O(2^{\ell})$. The complexity of the LDPC-aided check is given by $ O(\delta I)$, where $\delta$ represents the number of non-zero elements of the parity-check matrix ${\mathbf{H}}$. For a given $(\gamma,\rho)$-regular LDPC with code length $N$, we have $\delta = \gamma N$. Therefore, the complexity of the statistical checking for each candidate grows linearly with $N$. To summarize, the overall complexity of additional operations is $O(2^{\ell})+O(N I)$.

\section{Simulations}
For simulation, we select a rate-$1/2$ $(3,6)$-regular LDPC code $\mathcal{C}[2304,1152]$ which is constructed by the progressive-edge-growth algorithm. The sum-product algorithm with maximum 50 iterations is employed for the decoding of LDPC coded data.
In the following two different signal constellations are considered.

\begin{figure}[h]
  \centering
  \includegraphics[width=7cm]{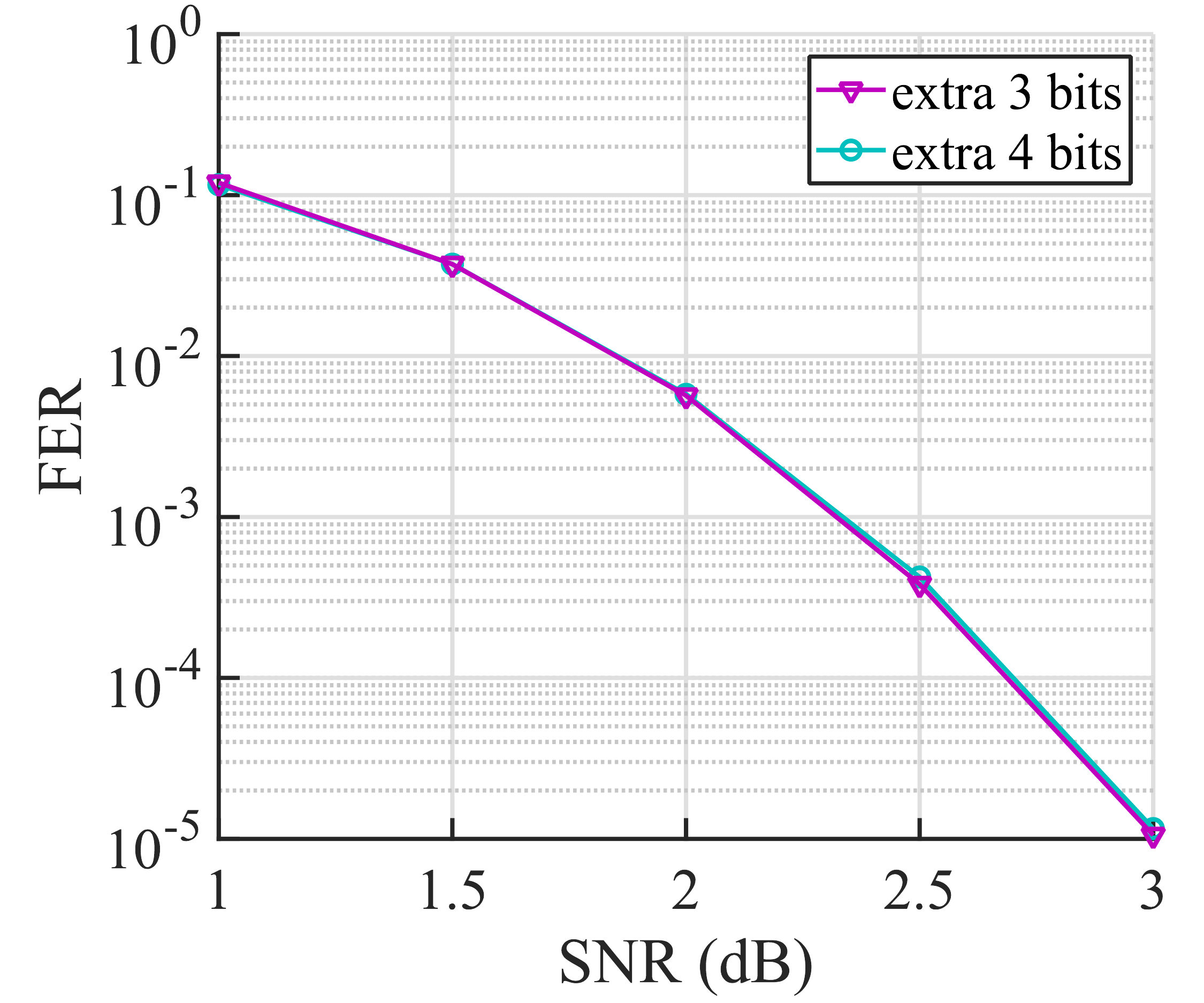}\\
  \caption{
  FER of the extra bits transmitted with the rotated QPSK constellation
  encoded by the LDPC code $\mathcal{C}[2304,1152]$.
  }\label{FIG8}
\end{figure}

\begin{figure}[h]
  \centering
  \includegraphics[width=7cm]{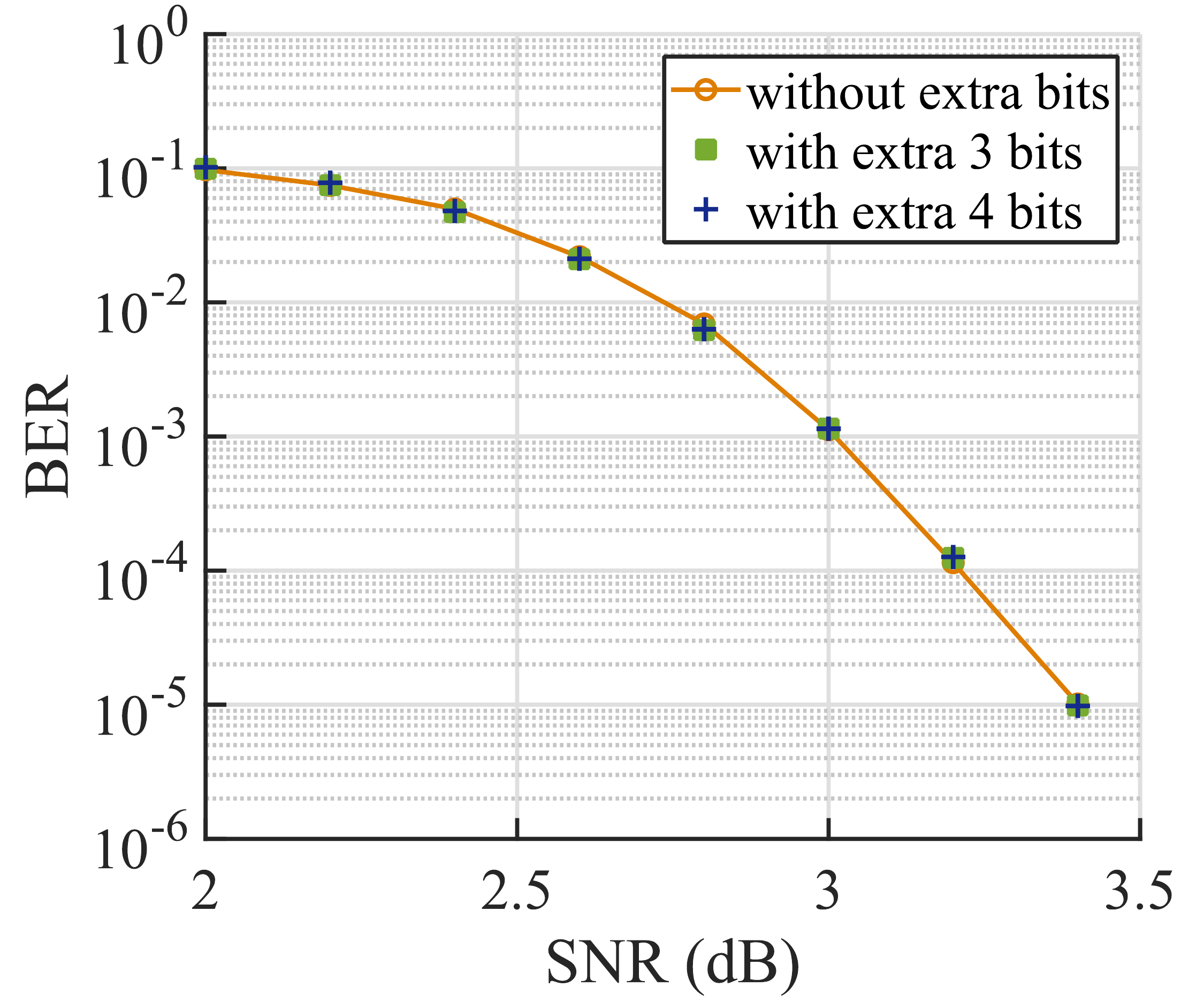}\\
  \caption{BER of the payload data encoded by the LDPC code $\mathcal{C}[2304,1152]$ with the rotated QPSK constellation.}\label{FIG7}
\end{figure}

\textbf{\emph{Example 2:}} In this example, we consider QPSK, where the permutation ambiguity occurs if the difference between two rotation angles is a multiple of ${\pi}/{2}$.  In particular, once we find a valid rotation angle $\widehat{\theta}$ by the brute-force search, the list of rotation angle candidates is given by $\Theta = \{\widehat{\theta},\widehat{\theta}+{\pi}/{2}, \widehat{\theta}+\pi,\widehat{\theta}+{3\pi}/{2}\}$, where ``$+$'' denotes the modulo $2\pi$ addition.
We investigate the frame error rate~(FER) of the extra bit sequence under the proposed scheme~(see Fig. \ref{FIG8}). It can be seen that, for instance, the FER with $\ell=4$ is about $10^{-5}$ at SNR of $3$ dB. Hence, the influence of erroneous decoding of the additional bits sequence on the error performance of the information sequence can be neglected, especially when the SNR is relatively large. To illustrate this, we present the bit error rate (BER) of the payload data under the proposed scheme, compared with that of the payload data without packing extra bits~(see Fig. \ref{FIG7}). As expected, the performances of the proposed scheme with $\ell=3$ and $\ell = 4$ are nearly the same as that without the transmission of additional bits.

\begin{figure}[h]
  \centering
  \includegraphics[width=7cm]{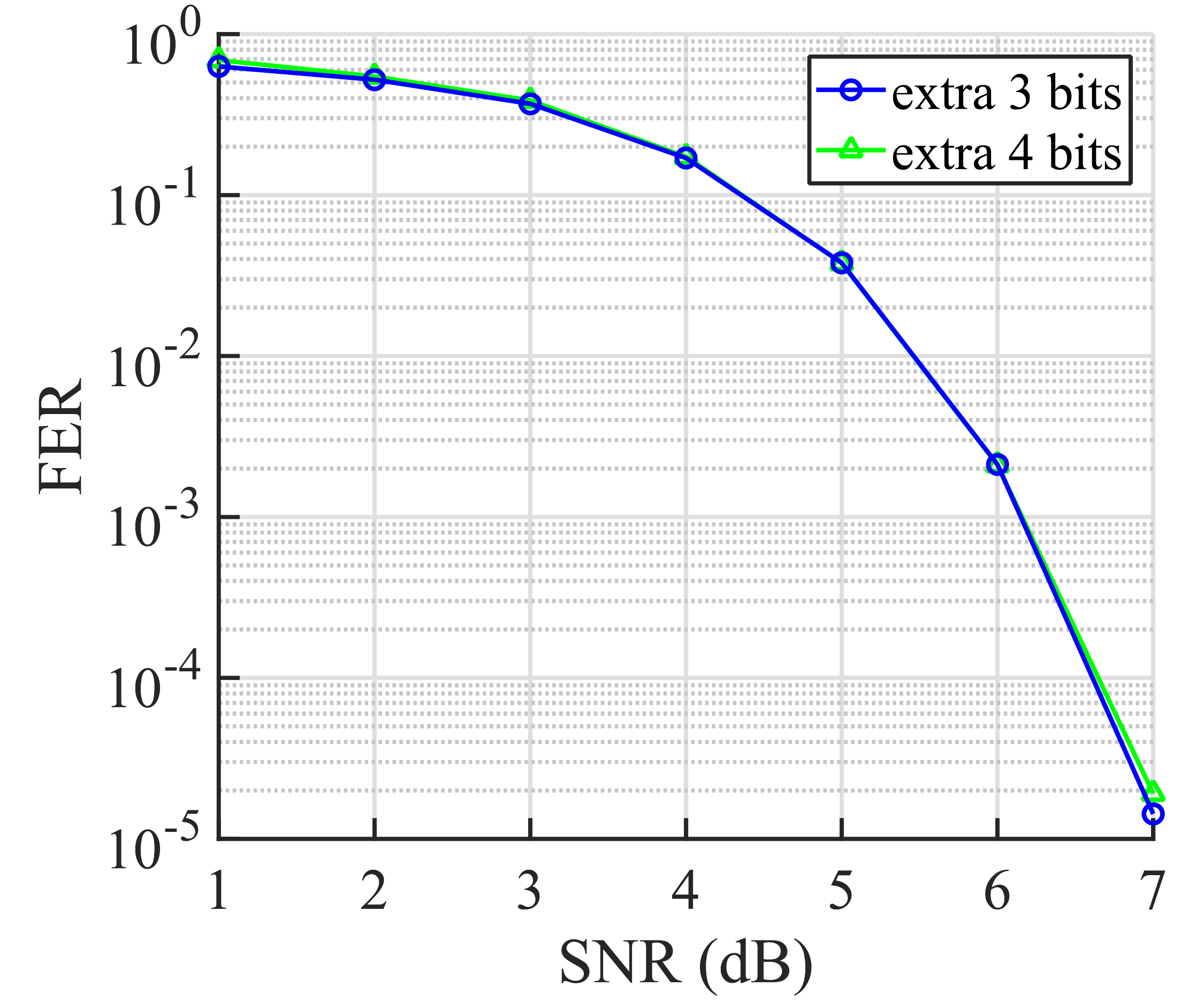}\\
  \caption{FER of the extra bits transmitted with the rotated 16QAM constellation
  encoded by the LDPC code $\mathcal{C}[2304,1152]$.}\label{FIG6}
\end{figure}

\begin{figure}[h]
  \centering
  \includegraphics[width=7cm]{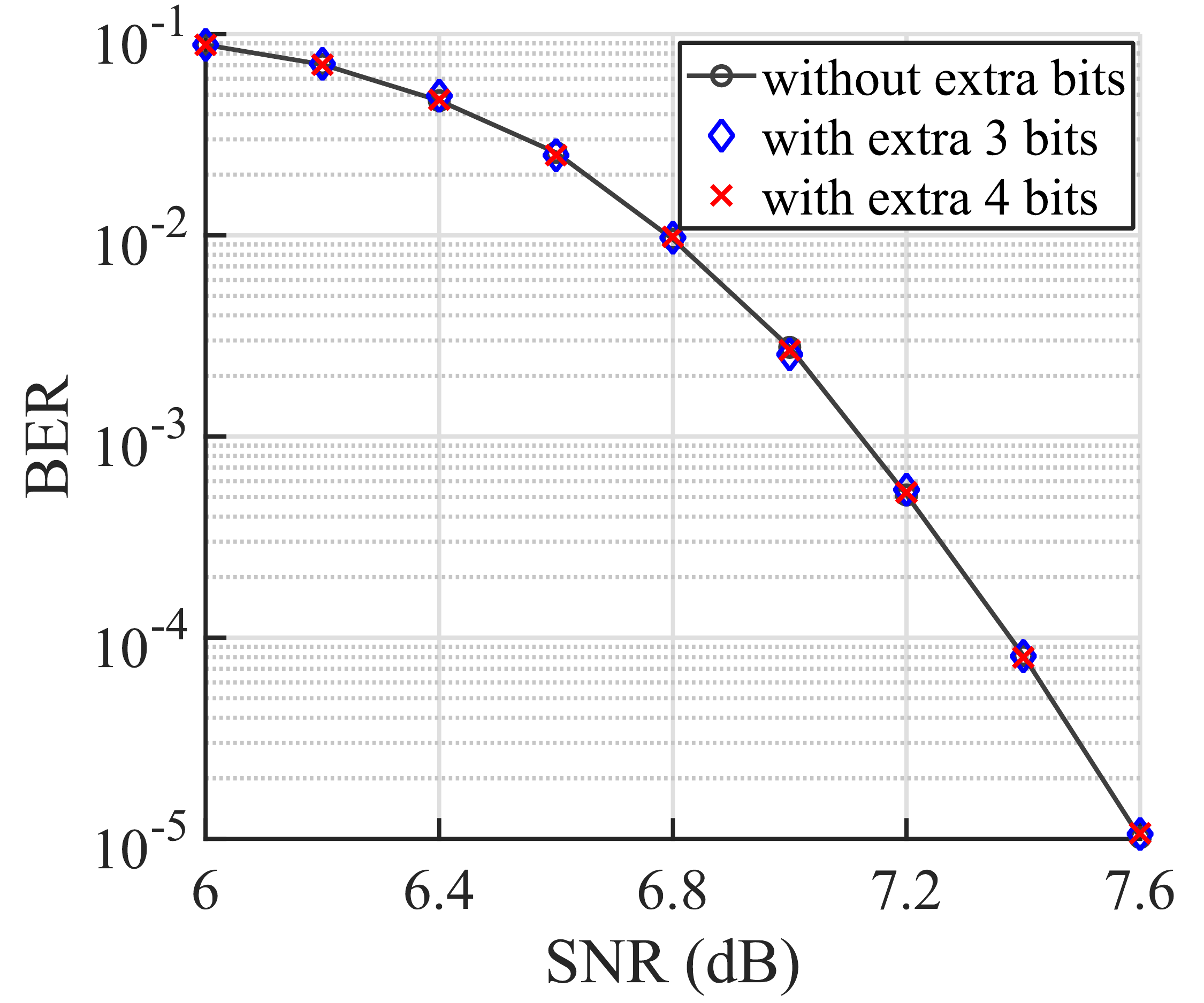}\\
  \caption{BER of the payload data encoded by the LDPC code $\mathcal{C}[2304,1152]$ with the rotated 16QAM constellation.}\label{FIG5}
\end{figure}

\textbf{\emph{Example 3:}} Consider the Gray 16QAM in  \textbf{\emph{Example 1}} again. Similar to QPSK, the list of rotation angle candidates on the 16QAM constellation is also given by $\Theta = \{\widehat{\theta},\widehat{\theta}+{\pi}/{2}, \widehat{\theta}+\pi,\widehat{\theta}+{3\pi}/{2}\}$.
We present the FER of the extra bits under the proposed scheme in Fig. \ref{FIG6}, where the erroneous decoding of the extra bits occurs in the relatively low SNR region. Nevertheless, this region is of little interest for the payload data as shown in Fig. \ref{FIG5}. One can see that the BER curves with $\ell=3,\ell=4$ overlap with that without packing extra bits.

\section{Conclusion}
In this letter, we have proposed to use rotated signal constellations combined with the LDPC code to achieve the transmission of extra bits. By embedding the extra bits into a rotation angle of the signal constellation, the extra bits can be transmitted without any increase in bandwidth or transmission power consumption. At the receiver, the rotation angle is estimated based on the brute-force search and the statistical learning of the syndrome of the LDPC code. Simulation results show that additional bits can be transmitted with negligible influence on the reliability of the LDPC coded data.

\vskip3pt
\ack{This work was supported by the Science and Technology Planning Project of Guangdong Province~(No. 2018B010114001), the NSF of China~(No. 91438101, 61771499) and the Basic Research Project of Guangdong Provincial NSF~(No. 2016A030308008).}

\vskip5pt

\noindent Jiachen Sun~(\textit{School of Electronics and Information Technology, Sun Yat-sen University, Guangzhou 510006, China})

\noindent Hao Liu and Xiao Ma~(\textit{School of Data and Computer Science, Sun Yat-sen University, Guangzhou 510006, China})
\vskip3pt

\noindent E-mail: maxiao@mail.sysu.edu.cn

\end{document}